\documentclass[aps,
twocolumn,
showpacs,
amsmath,
amssymb,
nofootinbib]{revtex4} 
\usepackage[dvips]{graphicx}

\begin{document}

\title{On existence of matter outside a static black hole}

\author{Tetsuya Shiromizu$^{(1)}$, Sumio Yamada$^{(2)}$ and Hirotaka Yoshino$^{(3,1)}$}


\affiliation{$^{(1)}$Department of Physics, Tokyo Institute of Technology, Tokyo 152-8551, Japan}

\affiliation{$^{(2)}$Mathematical Institute, Tohoku University, Sendai 980-8578, Japan}

\affiliation{$^{(3)}$Advanced Research Institute for Science and Engineering, 
Waseda University, Tokyo 169-8555, Japan}

\date{\today}

\begin{abstract}
It is expected that matter composed of a perfect fluid cannot be 
at rest outside of a black hole if the spacetime is asymptotically flat and 
static (non-rotating). However, there 
has not been a rigorous proof for this expectation without assuming spheical 
symmetry. In this paper, we provide 
a proof of non-existence of matter composed of  a perfect fluid in static 
black hole spacetimes under certain conditions, which can be interpreted as a relation 
between the stellar mass and the black hole mass. 
\end{abstract}


\maketitle

\section{Introduction}

The issue of the final state of gravitational collapse is important from various 
points of view. Usually we expect the black holes will eventually form  after
the gravitational collapse of a star and 
the spacetime 
is expected to asymptotically converge to a stationary or static state. 
The perturbation analysis \cite{Price} 
and numerical demonstrations support this picture. 
As a consequence, the limiting spacetime will be a stationary or 
static vacuum spacetime if all the matter is absorbed to 
the thus produced black hole. 
Nowadays we know that the uniqueness theorem for black holes holds in 
asymptotically flat, stationary (or static) and vacuum spacetimes and 
the resulting spacetimes are described by Kerr or Schwarzschild solutions 
\cite{Israel,4d3,Review}. 
Therefore we can have definite astrophysical predictions using these 
exact solutions.

While the uniqueness of the vacuum black hole has been established, 
it is interesting to ask if there is a uniqueness 
theorem for stationary or static spacetimes of a black hole plus matter. 
In the stationary case (i.e., time translation symmetry exists but their trajectories are 
not hypersurface orthogonal), clearly this is not the case because many 
spacetime solutions have been constructed. 
The perturbative solution of a slowly rotating black hole surrounded 
by an infinitesimal ring was constructed by Will~\cite{Will}. There are numerical solutions
of a black hole surrounded by an infinitely thin disk~\cite{Lanza} 
or by an differentially rotating ring~\cite{NE94}. Recently, 
the black hole with a uniformly rotating ring around it has been calculated 
with high accuracy~\cite{AP}. 

On the other hand, we expect that there should be some kind of uniqueness 
theorem for static spacetimes (i.e., trajectories of time-translation symmetry 
are hypersurface orthogonal) of a black hole with matter. 
This is because if we try to put the matter at rest outside a black hole, 
it is expected to fall into the black hole because of the gravitational attraction. 
However, we would like to point out that this intuitive picture 
does not have a rigorous reasoning. In fact, one can easily 
provide a counterexample as follows. If an infinitely thin disc composed 
of counter-rotating particles exists on the equatorial plane of the black hole spacetime, 
there should exist a static configuration after a fine tuning 
of system parameters. This example indicates 
that the uniqueness of spacetimes of a black hole with matter 
depends not only on the energy condition but also the equation of state 
matter satisfies.  As far as we know, there is Bekenstein's work \cite{Be} 
which proves the 
nonexistence of scalar fields outside a black hole in the static spacetime. 
However, without assuming the spherical symmetry, 
the question of nonexistence of matter composed of an
ordinary perfect fluid outside a black hole in the static spacetime has not been addressed  
up to now.  

As it turns out, proving the uniqueness in the  setting above 
is  rather a delicate problem. A similar situation 
is found for the proof of the spherical symmetry 
of a static isolated star composed of perfect fluid 
for a certain class of equations of state. 
This problem was  solved relatively recently 
by Lindblom and Masood-ul-Alam 
\cite{LM} after a long history \cite{M,BS}. 
In the work of \cite{LM}, the authors studied the condition 
that should hold inside of a star {\it without} assuming the 
spherical symmetry and  proved the 
conformal flatness of the space that directly implies the 
spherical symmetry in turn. 

Fortunately, many of the results in \cite{LM} can be used 
for our situation because much of their analysis 
was done in quite a general setting. In this paper 
we would like to reformulate  
the analysis of \cite{LM} 
for the goal of showing the uniqueness of the static spacetime of a black hole 
plus matter, taking care of the difference between 
the two setups. We will provide 
a partial proof for the non-existence of 
matter outside a static black hole. 
The non-existence we obtain here is conditional under an inequality between the black-hole mass 
and the stellar mass. 

This paper is organized as follows. 
In the next section, we provide our setup concerning the Einstein equation. 
In Sec.III, we summarize the results of Lindblom and Masood-ul-Alam~\cite{LM}, 
paying particular attention to the part that are directly relevant to our argument. 
We prove our theorem in Sec.IV and summarize our paper 
with some discussion in Sec.V. 
We adopt the unit of $c=G=1$. 

\section{Setup}

We consider the static spacetime which has the metric 
%
\begin{eqnarray}
ds^2=-V^2(x)dt^2+g_{ij}(x)dx^i dx^j,
\end{eqnarray}
%
where $i=1,2,3$ and $g_{ij}(x)$ is the induced metric of 
\{ $t=$ const.\} hypersurfaces $\Sigma$. In a static spacetime, its event horizon $H$ is identified
with Killing horizon \{$V=0$\} and thus $V(x)$ vanishes on the horizon. 
We assume that there is a perfect fluid 
with energy density $\rho$ and pressure $P$. 
The fluid is assumed to satisfy an equation of state $P=P(\rho)$. 
We assume the surface of the star/fluid is a two-dimensional 
closed connected  equipotential surface $\{x: V(x) = V_s > 0\}$ for some positive constant $V_s$.

\begin{figure}[bt]
\centering
{
\includegraphics[width=0.35\textwidth]{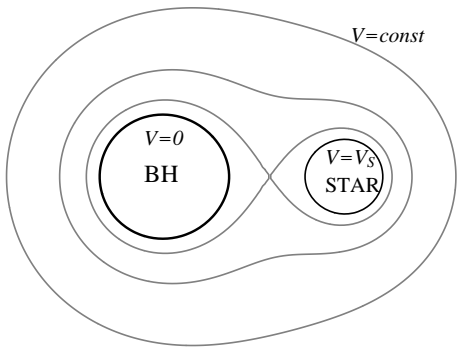}
}
\caption{Typical configuration of the system of a black hole (BH) and matter (STAR). 
}
\label{configuration}
\end{figure}
 
The Einstein equation and equation for fluid are given by 
%
\begin{eqnarray}
& & D^2 V =4\pi V (\rho + 3P), \label{00} \\ 
& & R_{ij}=\frac{1}{V}D_i D_j V + 4\pi (\rho -P)g_{ij}, \\ 
& & D_i P = -\frac{1}{V}(\rho +P) D_i V, \label{pf} 
\end{eqnarray} 
%
where $D_i$ and $R_{ij}$ are the covariant derivative and 
Ricci tensor of the metric $g_{ij}(x)$. 

Eq.~\eqref{pf} indicates that the surface $\{ \rho={\rm const.}\}$ 
is identical to the surface $\{ V={\rm const.} \}$ 
Let us suppose for the moment the condition $D_i V \neq 0$ at the horizon. Except for the 
extremal charged black holes, this condition is satisfied in all the known static 
black hole solutions. Then Eq.~\eqref{pf} 
says that the value of $D_iP$ diverges at the horizon 
if matter exists at the horizon $\{ V=0 \}$, which is an unphysical situation. 
Hence, this would make the stellar surface disjoint from the horizon. 
We stress that the same conclusion can be obtained from Eq. \eqref{00} without any 
additional hypothesis.  As Eq. \eqref{00} is elliptic, a standard boundary 
elliptic estimate (see, for example, Lemma 6.4 in Gilbarg-Trudinger \cite{GT}) says 
that near the horizon, the norm of the gradient of $V$ is 
bounded by the sup norm of the right-hand side of Eq. (\ref{00}) 
as well as that of $V$, both of which are uniformly bounded in our case.  
As the horizon is the zero set of $V$, while the surface of the star 
is the level set $\{V=V_s (> 0)\}$, the gradient bound for $V$ provides 
a positive lower bound (dependent on $V_s$) for the distance between those 
two level sets, which in turn implies that the star is disjoint from the horizon. 

At the moment  pictures such as Fig.1 are possible configurations of our 
setup.  Although the star surface shown in this 
figure is spherical, note that its topology is arbitrary as long as it is 
specified as a connected equipotential surface $\{V=V_s\}$. 
Thus our theorem will hold also for, e.g., a barotropic perfect fluid ring
\footnote{Note, however, that our theorem is not applicable to a ring composed
of counter-rotating particles around a black hole, as mentioned in Sec.I. 
This is because such matter has non-isotropic stress
and its equation of state is not barotropic.  }.

Here we stress that we will prove that such configurations as above do {\it not} occur 
under certain conditions.  In doing so, we are also {\it not} assuming any symmetry 
of the spacelike slice. 

Asymptotic flatness requires the following asymptotic behavior of $V$ and the metric 
%
\begin{eqnarray}
V = \left( 1-\frac{M}{r} \right) +O(r^{-2})
\end{eqnarray}
%
and
%
\begin{eqnarray}
g_{ij} = \left( 1+\frac{2M}{r} \right) \delta_{ij} +O(r^{-2}),
\end{eqnarray}
%
where $r=:|\delta_{ij}x^i x^j|^{1/2}$ and $M$ is the ADM mass. 

Our strategy for the proof is as follows.
We first show that the $\{ t= \mbox{const}.\}$ hypersurface $\Sigma$ is conformally 
flat. The main part of the proof is finding 
appropriate conformal transformations for $\Sigma$ so that it 
becomes a hypersurface with zero ADM mass and non-negative Ricci scalar 
curvature. This wonderful idea was first introduced by Bunting and Masood-ul-Alam \cite{BM}. 
We then apply the positive energy theorem \cite{SY} to 
this surface to conclude that the surface is flat Euclidean space. 
In turn, the original hypersurface $\Sigma$ is conformally flat and 
it will immediately follow that $\rho=P=0$. 
In order to find the appropriate conformal transformation, we will use 
the Lindblom and Masood-ul-Alam's theorems \cite{LM} that was used 
for proving the spherical symmetry of a static star. 
We review it in the next section. 

\section{Theorems of static stellar models}

In this section, we briefly review Lindblom and Masood-ul-Alam's results \cite{LM} 
where the spherical symmetry of static stellar models was obtained under a certain 
condition on the equation of state.  We are able to transcribe their argument mostly 
because the techniques used in treating the {\it inside} of stars are identical.

Because the $\{V={\rm const.}\}$ surface is identical to $\{\rho={\rm const.}\}$ 
surface due to Eq.~\eqref{pf} and the equation of state $P = P(\rho)$, 
both $\rho$ and $P$ are regarded as functions of $V$. 
The minimum value of $V$ in the star is denoted by $V=V_c$. 
Then quantities $r_\mu(V)$, $m_\mu(V)$ are defined as the solutions to the equations 
%
\begin{eqnarray}
\frac{dr_\mu}{dV}=\frac{r_\mu (r_\mu -2m_\mu)}{V(m_\mu+4\pi r_\mu^3P)},
\label{remark12}
\end{eqnarray}
%
%
\begin{eqnarray}
\frac{dm_\mu}{dV}=\frac{4\pi r_\mu^3 (r_\mu -2m_\mu)\rho}{V(m_\mu+4\pi r_\mu^3P)},
\label{remark13} 
\end{eqnarray}
%
where the boundary conditions for Eqs. (\ref{remark12}) and (\ref{remark13}) are 
%
\begin{eqnarray}
r_\mu (V_s)=R_\mu :=2\mu /(1-V_s^2), 
\end{eqnarray}
%
%
\begin{eqnarray}
m_\mu (V_s)=\mu. 
\end{eqnarray}
%
Here, $\mu$ 
is some constant suitably chosen in the proof. 
As we see from Eqs.~\eqref{remark12} and \eqref{remark13}, 
$dm_\mu/dr_\mu=4\pi r_\mu^2\rho$ holds and 
$\mu$ could be interpreted as the local mass of the star.
Now $W_\mu (V)$ is defined by 
%
\begin{eqnarray}
W_\mu := 
 \left\{ \begin{array} {ll} 
\displaystyle\frac{(1-V^2)^4}{16\mu^2} & \mbox{outside of the star}, \\ 
\displaystyle\left(1-\frac{2m_\mu}{r_\mu} \right)\left(\frac{dr_\mu}{dV} \right)^{-2}  
&\mbox{inside of the star.}  \end{array} 
\right. 
\label{remark11} 
\end{eqnarray}
%
Various lemmas on $r_\mu(V)$, $m_\mu(V)$, $W_\mu(V)$ are introduced, 
such as the existence of the solution on $(V_\mu, V_s]$ for some $V_\mu$ (Lemma 1-3). 
In particular, assuming that the pressure is finite $P(V)<\infty$, 
$W_\mu$ can be positive for somewhat small $\mu$ over the interval $(V_c, V_s]$ 
(Lemma 8).

We require that the equation of state satisfies at least one of the following constraints.
%
\begin{eqnarray}
\frac{1}{5}\kappa^2 +2\kappa +(\rho+P) \frac{d \kappa }{dP} \leq 0 \label{conA} 
\end{eqnarray}
%
or 
%
\begin{eqnarray}
\frac{5\rho^2}{6P(\rho+3P)} \geq \kappa > \frac{10V_s^2}{e^{2h(P)}-V_s^2}, \label{conB}
\end{eqnarray}
%
where $\kappa:=(d \rho/dP)(\rho+P)/(\rho+3P)$ and $h(P)={\rm log}(V_s/V)$. The lower 
bound on $\kappa$ in the condition (\ref{conB}) implies that the adiabatic index 
$\gamma:=(dP/d \rho)(\rho+P)/\rho$ must be less than or equal to $(6/5)(1+P/\rho)^2$. 
Under this assumption, the following quantity
%
\begin{eqnarray}
\Sigma_\mu := \frac{dW_\mu}{dV}-\frac{8\pi}{3}V(\rho+3P)
+\frac{4W_\mu}{5V}\frac{\rho+P}{\rho+3P}\frac{d \rho}{dP}. 
\end{eqnarray}
%
is proved to be non-negative (Lemma 10 \cite{LM}).

The most important technical result is Lemma 6 in \cite{LM}:
{\it Assuming that $W_\mu >0$ and $\Sigma_\mu \geq 0$ on 
$(V_c,V_s]$, 
then $W_\mu \geq W:=D_a V D^a V$ on $(V_c,V_s)$
holds everywhere} \footnote{In  Lemma 6 of Ref. \cite{LM}, 
it is stated that $W_\mu \geq W$ holds on $(V_c,V_s]$. However, we can 
easily see from the proof 
that it holds everywhere outside of the star.
} 
{\it and $\mu \leq M$}. 
Because this lemma is relevant to our problem, we briefly review its proof. 
In \cite{LM}, a differential inequality of elliptic type in the exterior vacuum region is derived: 
%
\begin{eqnarray}
D_a(V^{-1}D^a \Delta^+_\mu) \geq 0, 
\end{eqnarray}
%
where 
%
\begin{eqnarray}
\Delta^+_\mu := (W-W_\mu) \frac{(1-V_s^2)^3(1+b-V_s^2)}{(1-V^2)^3(1+b-V^2)}.
\end{eqnarray}
%
$b$ is some positive constant to be chosen later. 
The appropriate boundary condition at infinity is $W-W_\mu = 0$. 
From the maximum principle, $\Delta^+_\mu$ has the maximum value at 
the infinity or at the surface of the star $ \{ V=V_s \}$. 
The possibility of maximum value of 
$W-W_\mu$ occurring at $\{V=V_s\}$ is then removed by choosing $b$ appropriately. 
Thus $\Delta^+_\mu \leq 0 $ holds everywhere 
and $W_\mu \geq W$. 
$\mu \leq M$ is directly derived from the asymptotic 
expansion of  $W_\mu \geq W$. 
Note here that in our case where the black hole is additionally present, 
the possibility that $\Delta^+_\mu $ takes a maximal value
at the event 
horizon cannot be removed, and we cannot use Lemma 6 in \cite{LM}
without modifications.  We will come back to this point in the next section.

They next introduce $M^-$ and $M^+$ 
where $M^-$ satisfies $M^- < M$ and $W_{\mu=M^-}>0$ on $[V_c,V_s]$ 
while $M^+$ satisfies $M^+ >M$ with $W_{\mu=M^+}(V_{M^+_*})=0$ 
at a point $V_{M^+_*} \in (V_c,V_s)$. 
The existence of $M^\pm$ is guaranteed by, e.g., Lemma 6~\cite{LM}. 
Then, they defined $\nu$ by 
$\nu =\inf_{\mu \in  S_c} \mu$ where 
$S_c= \lbrace \mu \in [M^-,M^+]:W_\mu >0~{\rm on}~(V_c,V_s]~{\rm and}~W_\mu(V_c)=0 
\rbrace$.

In their Lemma 14, they proved that 
$d^2\psi_\nu/dV^2$ is positive, 
where $\psi_\mu(V)$ is defined as the solution to the equation 
%
\begin{eqnarray}
\frac{d \psi_\mu}{dV}=\frac{\psi_\mu}{2r_\mu {\sqrt {W_\mu}}} 
\left(1-\frac{2m_\mu}{r_\mu} \right), 
\end{eqnarray}
%
with $\psi_\mu(V_s)=(1+V_s)/2$ at the surface of the star $\{V=V_s \}$.
Finally they define a conformal metric $g^+_{ij}=\Omega_+^4g_{ij}$, 
where
\begin{eqnarray}
\Omega_+ = \left\{ \begin{array} {ll} 
(1+V)/2 & \mbox{ outside of the star,} \\ 
\psi_\nu(V) &\mbox{ inside of the star. }  \end{array} 
\right. \label{omega+} 
\end{eqnarray} 
and found that 
the Ricci scalar has the expression 
%
\begin{eqnarray} 
R(g_{ij}^+)=(\tilde W-W) \frac{8}{\Omega^5_+} \frac{d^2 \Omega_+}{dV^2}, \label{ricci+} 
\end{eqnarray} 
%
where $\tilde W=W_\nu$ \footnote{ 
In defining $\tilde W$ precisely,  
we must take care of the differential structure 
at the center of stars. As the argument is rather 
technical, we refer this point to 
{\it Lemma 11, 12, 13, 14} and the main theorem in Ref. 
\cite{LM}.}. 
By Lemma 6 and 14, this is positive, while the space has zero ADM mass. 
Here the case of equality in the positive energy theorem can be applied and we
conclude that the space is conformally flat, which then implies that an isolated star 
is spherically symmetric \cite{Lee}.

\section{Proof of non-existence of matter outside a static black hole}

Now we turn our attention to the proof of non-existence of matter 
outside a static black hole. 
The various functions {\it inside} of a star, 
$r_\mu$, $m_\mu$, $\psi_\mu$, $W_\mu$ and $\Delta_\mu^+$ 
can be introduced without modifications. 
We adopt the same condition on the equation of state. 
However, the modification for Lemma 6 is required as 
mentioned in the previous section. 

Lemma 6 in \cite{LM} says that $\tilde W-W \geq 0$ 
holds everywhere. 
In the case of \cite{LM}, this inequality was proved by 
showing $\Delta_\mu^+<0$ at $V=V_s$ and by 
applying the maximum 
principle for the inequality $D_a(V^{-1}D^a\Delta_\mu^+)\ge 0$ which indicates 
that $\Delta_\mu^+=0$ at infinity is the maximum value. 
However, presently we have an additional boundary, 
which is the horizon $\{V=0\}$, and a possibility of 
$\Delta_\mu^+>0$ there remains.  We thus have to 
{\it assume} $\Delta_\mu^+<0$ at the horizon or equivalently 
%
\begin{eqnarray} 
(W-W_\mu)_{V=0} \leq 0. \label{una} 
\end{eqnarray} 
%
We will discuss the physical meaning of this hypothesis in the last section. 
Under this assumption, $W_\mu-W \geq 0$ is guaranteed everywhere, 
and the results in \cite{LM} that was proven using Lemma 6 
become available. In particular, the existence of $\nu$ is guaranteed. 

Now we show the conformal flatness of this space. 
Consider the two conformal transformations defined by 
%
\begin{eqnarray}
g_{ij}^\pm = \Omega_\pm^4 g_{ij}, 
\end{eqnarray}
%
where $\Omega_+$ is the same as Eq.~\eqref{omega+} and
%
\begin{eqnarray}
\Omega_- = \frac{1}{2}(1-V). 
\end{eqnarray}
%
Now we have two manifolds $(\Sigma^\pm, g_{ij}^\pm)$. 
As in Bunting and Masood-ul-Alam \cite{BM}, we can make a smooth manifold 
out of $\Sigma^+ \cup \Sigma^-$ by gluing along  $\{V=0\}$.

The asymptotic behavior of each metric is; 
%
\begin{eqnarray}
g_{ij}^+=\delta_{ij}+O(r^{-2}),
\end{eqnarray}
%
and
%
\begin{eqnarray}
g_{ij}^- dx^i dx^j = (M/2r)^4 (dr^2+r^2d \Omega^2_2) +O(r^{-5}).
\end{eqnarray}
%
In the manifold $(\Sigma^-, g_{ij}^- )$, $\{r = \infty \}$ corresponds 
to a regular point in a 3-dimensional surface. Indeed, introducing a new coordinate 
$R=M^2/4r$, the metric near $r=\infty$ can be written as 
%
\begin{eqnarray}
g_{ij}^- dx^i dx^j = dR^2+R^2d \Omega^2_2. 
\end{eqnarray}
%
If the Ricci scalar of this space is non-negative, 
positive energy theorem tells us that $\Sigma^+ \cup \Sigma^-$ is 
the flat Euclidean space and thus $\Sigma$ is conformally flat, 
because this manifold has the zero ADM mass and non-negative Ricci scalar. 
Hence we want to 
show the non-negativity of the Ricci scalars $R(g^\pm_{ij})$.

The three-dimensional 
Ricci scalar becomes the same as \eqref{ricci+} for $\Sigma^+$ and 
%
\begin{eqnarray} 
R(g_{ij}^-)=8\pi \Omega_-^{-5}\Bigl[(1+V) \rho + 6 VP \Bigr] \geq 0,
\label{ricci-}
\end{eqnarray}
%
for $\Sigma^-$. 
Since $R(g^-_{ij}) \geq 0$ holds as above, 
it remains to show the non-negativity of $R(g_{ij}^+)$. 
$d^2\Omega_+/dV^2\ge 0$ is guaranteed by Lemma 14 in \cite{LM} 
and $(\tilde{W}-W)\geq0$ is guaranteed by the hypothesis \eqref{una}. 
Then the conformally transformed space $\Sigma^+ \cup \Sigma^-$ 
has non-negative Ricci scalar (See Eq. (\ref{ricci+})) 
and zero ADM mass, which implies that $\Sigma^+\cup\Sigma^-$ is flat. 

Now we have proven under the assumption \eqref{una} 
that the original space $\Sigma$ is conformally flat. 
Going back to Eqs.~\eqref{ricci+} and \eqref{ricci-},  
we find that $R(g^{+}_{ij})=R(g^{-}_{ij})=0$ holds and 
they in turn imply 
$\tilde{W}=W$ and $\rho=P=0$, respectively. 
The latter implies that the spacetime should be vacuum. 
Our result excludes any static configurations of a black hole 
with a star whose surface is given by $\{V=V_s\}$ as was 
suggested in Fig.1.

We summarize  what we have obtained as the following theorem: 
\medskip 
\\ 
{\bf Theorem:} {\it In asymptotically flat static black hole spacetimes, 
the star, which is composed of a perfect fluid satisfying the dominant 
energy condition and has the surface of level surface set 
$\lbrace V=V_s (>0) \rbrace $, cannot 
exist if (i) the equations of state $P=P(\rho)$ 
satisfies the condition \eqref{conA} or 
\eqref{conB}, and 
(ii) for $W_\mu$ defined by Eq. \eqref{remark11} with Eqs. \eqref{remark12} and 
\eqref{remark13}, the inequality $W-W_\mu \le 0$ holds on the 
event horizon ($V=0$). }
\medskip

As a result, the ordinary uniqueness theorem~\cite{Israel} of vacuum spacetimes 
can be applied and then the spacetime in our setup is reduced to the 
Schwarzschild spacetime.

\section{Discussion}

We have proven the non-existence of matter composed of a perfect fluid 
under the hypothesis $(W-W_\mu)_{V=0} \leq 0$. 
Because the physical meaning of this hypothesis is still unclear, 
we examine it in this section. 
On the event horizon, the value of $\sqrt{W}$ coincides 
with the surface gravity $\kappa_{H}$ of the black hole. 
We introduce the Komar integral on the event horizon 
$M_{\rm BH}=\kappa_{H} A_H/4\pi$ that indicates the 
local mass of the black hole, where $A_H$ is the area of the horizon. 
Then, $(W-W_\mu)_{V=0}\le 0$ corresponds to 
\begin{equation} 
\mu\le \frac{1}{4\kappa_{H}}=\frac{A_H}{16\pi M_{\rm BH}}. 
\label{rewrote-condition} 
\end{equation} 
The first law of the black hole thermodynamics for the static spacetime 
$\delta M_{\rm BH}=\kappa_H\delta A_H/8\pi$ 
implies $A_H \propto M_{\rm BH}^2$. Hence  $A_H \sim 16\pi M_{\rm BH}^2$ 
and the right hand side of the inequality \eqref{rewrote-condition} 
is $O(M_{\rm BH})$.  
In order to find some upper bound on $\mu$ in terms of the 
quantities of the star, 
recall that $(W-W_\mu)_{V=V_s}\le 0$ holds on the surface of the star 
as appeared in the proof of Lemma 6 of \cite{LM}. This is rewritten as 
$\sqrt{W_s}\le (1-V_s^2)^2/4\mu$. 
Integrating over the surface of the star, we find 
\begin{equation}
\mu\le \frac{(1-V_s^2)^2}{16\pi M_\star}A_s, 
\end{equation} 
where $M_\star$ is the Komar mass of the star and 
$A_s$ is the area of the surface of the star. Hence, 
$(W-W_\mu)_{V=0}\le 0$ holds if  
\begin{equation}
(1-V_s^2)^2\frac{A_s}{16\pi M_\star}\le \frac{A_H}{16\pi M_{\rm BH}}\sim {M_{\rm BH}}.
\end{equation}
In order to simplify this inequality, 
let us consider the situation where 
a ball-shaped star exists outside a black hole and 
the distance between them is sufficiently large. 
In this case $V_s^2$ is approximated as $V_s^2\simeq 1-2M_\star/r_s$ 
and $A_s\simeq 4\pi r_s^2$, where $r_s$ is the radius of the star. 
Then the inequality becomes 
\begin{equation} 
M_\star\lesssim M_{\rm BH}. 
\end{equation} 
Therefore the main theorem states that a star with smaller mass than 
a black hole mass cannot exist outside of the black hole 
in static spacetimes.

Intuitively, heavy stars are also not permitted  
to exist outside of black holes in static spacetimes. 
We hope to have a different argument for proving the statement 
because our current 
approach of adapting 
Lindblom and Masood-ul-Alam's results \cite{LM} is not expected 
to produce optimal statements. 

Lastly we note that the star surface is assumed to be a surface of one 
connected component. In order to exclude 
the configuration of a star whose surface has two or more components 
such as a shell surrounding the black hole, further considerations are needed.


\section*{Acknowledgements}

The work of TS was supported by Grant-in-Aid for Scientific 
Research from Ministry of Education, Science, Sports and Culture of 
Japan(No.13135208, No.14102004, No. 17740136 and No. 17340075), 
the Japan-U.K. and Japan-France Research  Cooperative Program. 
The work of SY is supported by Grant-in-Aid for Scientific 
Research (No.17740030,) as well as  Exploratory Research Program 
for Young Scientists from Tohoku University. 
The work of HY was partially supported by a Grant for The 21st Century 
COE Program (Holistic Research and Education Center for Physics 
Self-Organization Systems) at Waseda University.


\begin{thebibliography}{22}


\bibitem{Price}
R. H. Price, Phys. Rev. {\bf D5}, 2419(1972); ibid {\bf D5}, 2439(1972). 


\bibitem{Israel}
W. Israel, Phys. Rev. {\bf 164}, 1776(1967). 

\bibitem{4d3}
B. Carter, Phys. Rev. Lett. {\bf 26}, 331 (1971);
S.W. Hawking, Commun. Math. Phys. {\bf 25}, 152 (1972);
D.C. Robinson, Phys. Rev. Lett. {\bf 34}, 905 (1975).

\bibitem{Review}
B. Carter, 
and Black Hole Configurations.''
in {\it Gravitation in Astrophysics},
edited by B.~Carter and J.B. Hartle (Plenum, New York, 1987);
M.~Heusler, {\it Black Hole Uniqueness Theorems},
(Cambridge Univ. Press, London, 1996).

\bibitem{Will} C.M. Will, Astrophys. J. {\bf 191}, 521 (1974); {\bf 196}, 41 (1975).

\bibitem{Lanza} A. Lanza, Astrophys. J. {\bf 389}, 141 (1992).

\bibitem{NE94} S. Nishida and Y. Eriguchi, Astrophys. J. {\bf 427}, 429 (1994).

\bibitem{AP}
M.~Ansorg and D.~Petroff,
Phys.\ Rev.\ D {\bf 72}, 024019 (2005)
[arXiv:gr-qc/0505060];
D.~Petroff and M.~Ansorg,
arXiv:gr-qc/0511102;
M.~Ansorg and D.~Petroff,
arXiv:gr-qc/0607091.


\bibitem{Be}
J. D. Bekenstein, Phys. Rev. {\bf D5}, 1239(1972); 2403(1972). 

\bibitem{GT} D. Gilbarg and N. Trudinger, {\it Elliptic partial differential equations of second order},
(Springer, 1998)


\bibitem{LM}
L. Lindblom and A. K. M. Masood-ul-Alam, Commun. Math. Phys. {\bf 162}, 123(1994). 

\bibitem{M}
A. K. Masood-ul-Alam, Class. Quantum Grav. {\bf 5}, 409(1988). 

\bibitem{BS}
R. Beig and W. Simon, Commun. Math. Phys. {\bf 144}, 373(1992). 

\bibitem{BM}
G. L. Bunting and A. K. M. Masood-ul Alam, Gen. Rel. Gravi. {\bf 19}, 147(1987).

\bibitem{SY}
R. Schoen and S.-T.Yau, Comm. Math. Phys. {\bf 65}, 45(1979); ibid {\bf 79}, 231(1981). 

\bibitem{Lee}
L. Lindblom, J. Math. Phys. {\bf 21}, 1455(1980);
A. Avez, Ann. Inst. Henri Poincare {\bf A1}, 291(1964);
H. P. Kunze, Commun. Math. Phys. {\bf 20}, 85(1971). 

\end{thebibliography}
\end{document}